\documentclass[aps,prb,twocolumn,showpacs,superscriptaddress]{revtex4-1}\begin{tiny}\end{tiny}
\usepackage{longtable}
\usepackage{amsfonts}
\usepackage{graphicx}
\usepackage{epsfig}
\usepackage{setspace}
\usepackage{mathtools}
\DeclareMathOperator{\tr}{tr}

\begin{document}

\title{Recovering hidden Bloch character: Unfolding Electrons, Phonons, and Slabs}

\author{P. B. Allen}
\email{philip.allen@stonybrook.edu}
\affiliation{Department of Physics and Astronomy, Stony Brook University, Stony Brook, NY 11794-3800, USA}

\author{T. Berlijn}
\affiliation{Physics Department, University of Florida, Gainesville, FL 32611-8440, USA}

\author{D. A. Casavant}
\affiliation{Department of Physics, University of Maryland, College Park, MD 20742-4111, USA}

\author{J. M. Soler}
\affiliation{Dep. de F\'isica de la Materia Condensada, Universidad Aut\'onoma de Madrid, 28049 Madrid, Spain}

\date{\today}


\begin{abstract}
For a quantum state, or classical harmonic normal mode, of a system of spatial periodicity ``$R$,'' 
Bloch character is encoded in a wavevector ``$K$.''  
One  can ask whether this state has partial Bloch character ``$k$'' corresponding to a finer scale of periodicity ``$r$.''  
Answering this is called ``unfolding.''  A theorem is proven that yields a mathematically clear prescription
for unfolding, by examining translational properties of the state, requiring no ``reference states'' or basis functions
with the finer periodicity ($r,k$).  A question then arises, how should one assign partial Bloch character to a state of a
finite system?  A slab, finite in one direction, is used as the example.  Perpendicular components $k_z$ 
of the wavevector are not explicitly defined,
but may be hidden in the state (and eigenvector $|i\rangle$).  A prescription for extracting $k_z$ is offered and tested.
An idealized silicon (111) surface is used as the example.  Slab-unfolding reveals
 surface-localized states and resonances which were not evident from dispersion curves alone.
 \end{abstract}

\maketitle

 
\section{Unfolding in General}
\label{sec:general}
 
An alloy, and various other complicated systems, may ``inherit'' approximate 
translational order from an underlying simpler periodic system.  For example,
``modulated crystals'' \cite{Wiegers,Vansmaalen} have a double-layer structure, 
each layer with its own ``inherent'' two-dimensional translational symmetry. 
Adjacent layers have translations incommensurable with each other.  Interlayer interaction modulates both translations.
 Computations are done in a supercell,
chosen to be as close a multiple as is feasible of the two different inherent translations.
The eigenstates of the system remember, in some approximation, 
the periodicity of one, or the other, or both, of the unmodulated lattices.
One should then ask, for a given supercell eigenstate $|i\rangle$, how closely does it correspond to 
either of the two inherited families of Bloch states $|\vec{k}\rangle$?
``Closeness'' is measured by weights $W_i(\vec{k})$ between 0 and 1, with 0 and 1 meaning no inheritance 
or full.  Dispersions can then be plotted in an extended Brillouin zone of a simpler system.  Definitions
of unfolding algorithms were given Ku, Berlijn, and Lee (KBL) {\it et al.}\cite{Ku}, and by Popescu
and Zunger \cite{Popescu}.  Many earlier papers have used related versions of unfolding \cite{Baroni,Giustino}.
A new application to alloys contains interesting views. \cite{Haverkort}
The discussion by Popescu and Zunger\cite{Popescu} provides a convenient notation.

A simple type of unfolding has been used in discussing eigenstates of multilayers \cite{Jusserand}.  For example,
Scamarcio {\it et al.} \cite{Scarmacio} compute phonons in (Si)$_m$(GaAs)$_n$ multilayers, and relate normal modes confined
in the silicon layers to vibrational normal modes unfolded into a silicon crystal lattice.  A more closely related application
(but with weights always 0 or 1) is unfolding optical phonons measured in SiC polytypes into a larger Brillouin zone of a simpler
structure of SiC.  This was done by Feldman {\it et al.} \cite{Feldman} and Karch {\it et al.} \cite{Karch}.

The aim of this paper is to generalize the definition of unfolding.  Finite systems pose a particular challenge.  
A recipe for unfolding slab calculations is proposed.
  
\subsection{Notational preliminaries}
\label{sec:notation}

Assume that complicated system eigenstates are available from computation, 
performed with periodic boundary conditions.  The ``supercell'' (SC)  has translation 
generators $\vec{A}_i \ ( i=1,\ldots,3)$.
These are usually integer multiples of ``primitive cell'' (PC) translation generators $\vec{a}_i \ ( i=1,\ldots,3)$.  Upper 
and lower case will be used for SC and PC properties.  The integer multiple relation is 
$\vec{A}_i=\sum_j N_{ij} \vec{a}_j$, where
the $3\times 3$ matrix $\hat{N}$ has integer entries.
Vectors $\vec{R}$ will always be SC translations $\vec{R}=\sum_i m_i \vec{A}_i$, vectors $\vec{r}$ always
PC translations $\vec{r}=\sum_i n_i \vec{a}_i$, and when needed, the symbol $\vec{x}$ will denote an
arbitrary location in space.

The SC eigenstates $|\vec{K}J\rangle$ have the $\vec{K}$-space translational symmetry 
$\hat{T}(\vec{R})|\vec{K}J\rangle=\exp(i\vec{K}\cdot\vec{R})|\vec{K}J\rangle$, for SC translations 
${\hat T}$ by distance $\vec{R}$.
The Bloch vectors $\vec{K}$ lie in the supercell Brillouin Zone (SBZ) which is the unit cell of the SC reciprocal lattice.
Its translation generators are $\vec{B}_i$, where $\vec{A}_i \cdot \vec{B}_j=2\pi\delta_{ij}$.  
Reciprocal lattice translation generators $\vec{B}_i$ of the SC and $\vec{b}_j$ of the PC are related by
the inverse of the transpose of the integer matrix $N_{ij}$, specifically, 
$\vec{B}_i=\sum_j ((N^T)^{-1})_{ij} \vec{b}_j$.  Vectors $\vec{G}$ will always be SC reciprocal space
translations $\vec{G}=\sum_i m_i \vec{B}_i$, and vectors $\vec{g}=\sum_i n_i \vec{b}_i$ will be
PC reciprocal space translations.

The states $|\vec{K}J\rangle$ are not expected to have $\vec{k}$-space translational
symmetry: $\hat{T}(\vec{r})|\vec{K}J\rangle \ne \exp(i\phi)|\vec{K}J\rangle$, if $\vec{r}$ is a translation of the
PC but not of the SC.  However,
suppose the SC is nothing other than an exact ${\cal N}$-fold repetition of the PC 
(where ${\cal N} = \det{\hat{N}}$).  Then the SC has hidden
translational symmetry, and it would be possible to choose its eigenstates to have the Bloch property
$\phi=\vec{k}\cdot\vec{r}$, where the Bloch vector $\vec{k}$ would lie in the primitive cell Brillouin zone (PBZ),
being related to the supercell Bloch vector $\vec{K}$ by $\vec{k}=\vec{K}+\vec{G}$, for some SC reciprocal
lattice vector $\vec{G}$.  In this case, unfolding would be an exact simplification.

The volumes $\omega$ and $\Omega$ of the PC and SC are related by $\Omega/\omega=\det{\hat {\sf N}} \equiv {\cal N}$.
There are exactly ${\cal N}$ distinct PC translations by multiples of $\vec{a}_i$ that generate the SC from the PC.  These will
be labeled as $\vec{r}_i, \ i=1,\ldots,{\cal N}$.  There are similarly exactly ${\cal N}$ reciprocal space translations by multiples of $\vec{B}_i$ 
that generate the PBZ from the SBZ.  These will be labeled as $\vec{G}_i, \ i=1,\ldots,{\cal N}$.  These conjugate sets of 
discrete real and reciprocal space translations obey the ``Fourier'' relations
\begin{equation}
\frac{1}{{\cal N}}\sum_{i=1}^{\cal N}  e^{i\vec{G}_i \cdot \vec{r}_j}=\delta(j,0),
\label{eq:Sum1}
\end{equation}
\begin{equation}
\frac{1}{{\cal N}}\sum_{k=1}^{\cal N}  e^{i(\vec{G}_i - \vec{G}_j) \cdot \vec{r}_k}=\delta(i,j).
\label{eq:Sum2}
\end{equation}
These are discrete versions of the familiar ``quasi-continuum'' Fourier relations of the $\vec{k}$-vectors of a 
normal Brillouin zone (BZ)
and the discrete translations of a normal crystal.  The crystal's translation group was made finite
by the mechanism of  Born-von Karman periodic boundary conditions.  One way to 
understand Eq.(\ref{eq:Sum2}) is as a group representation orthogonality relation.  The discrete translations $\vec{r}_i$ indeed form a
finite Abelian translation group under the closure that obtains when the supercell translations are regarded as identity operators.

\subsection{unfolding theorem}
\label{sec:theorem}

Suppose we have a function $\Psi_{\vec{K}}$ with Bloch symmetry of the supercell: $\Psi_{\vec{K}}(\vec{x}+\vec{R}) 
=\exp(i\vec{K}\cdot\vec{R})\Psi_{\vec{K}}(\vec{x})$.  The Bloch wavevector is an arbitrary point 
$\vec{K}$ in the SBZ.  This section shows that there is a unique decomposition of this function, 
$\Psi_{\vec{K}}=\sum_G \psi_{\vec{K}+\vec{G}}$
into ${\cal N}$ functions having the additional Bloch symmetry of a primitive cell, $\psi_{\vec{k}}(\vec{x}+\vec{r})
=\exp(i\vec{k}\cdot\vec{r})\psi_{\vec{k}}(\vec{x})$, where $\vec{k}=\vec{K}+\vec{G}$
and $\vec{G}$ is any of the ${\cal N}$ $\vec{G}_i$'s.  The wavevectors $\vec{K}+\vec{G}$
lie in the PBZ, but only the $\vec{G}=0$ part lies in the SBZ.  The starting function $\Psi_{\vec{K}}$ is normalized to 1, but the
partial functions $\psi_{\vec{K}+\vec{G}}$ are not.  Then a candidate invariant weight for unfolding is the norm of the partial
function, $W_{\vec{K}}(\vec{G})=\int d^3\vec{x}|\psi_{\vec{K}+\vec{G}}(\vec{x})|^2$.  
Note that the partial functions obtained by unfolding are not in general eigenfunctions of any particular Hamiltonian.
The function being unfolded, $\Psi_{\vec{K}}$, need not be an eigenfunction; it only needs Bloch symmetry.
The translations $\vec{r}$ of the PC need not have a meaningful crystallographic relation to the 
translations $\vec{R}$ of the SC, except for having some integer commensurability relation $\vec{A}_i=\sum_j N_{ij} \vec{a}_j$.

Start by defining an operator $\hat{P}(\vec{K}\rightarrow\vec{K}+\vec{G})$, that operates on the SC Bloch
function $\Psi_{\vec{K}}$, and projects out the component that has the additional Bloch symmetry $\vec{K}+\vec{G}$.
\begin{equation}
\hat{P}(\vec{K}\rightarrow\vec{K}+\vec{G})=\frac{1}{{\cal N}}\sum_{i=1}^{\cal N} \hat{T}(\vec{r}_i)e^{-i(\vec{K}+\vec{G})\cdot\vec{r}_i}
\label{eq:P}
\end{equation}
where $\hat{T}(\vec{r}_i)f(\vec{x})=f(\vec{x}+\vec{r}_i)$.
To prove the projective property, translate by a PC translation $\vec{r}_j$ the function obtained by operating with $\hat{P}$
on a test function $\Psi_{\vec{K}}$.
\begin{eqnarray}
&&\hat{T}(\vec{r}_j)\hat{P}(\vec{K}\rightarrow\vec{K}+\vec{G})\Psi_{\vec{K}}(\vec{r}) \nonumber \\
&=&\frac{e^{i(\vec{K}+\vec{G}) \cdot \vec{r}_j}}{{\cal N}} \sum_{i=1}^{\cal N} \hat{T}(\vec{r}_i+\vec{r}_j) 
e^{-i(\vec{K}+\vec{G}) \cdot (\vec{r}_i+\vec{r}_j)}
\Psi_{\vec{K}}(\vec{r}) \nonumber \\
&=&e^{i(\vec{K}+\vec{G}) \cdot \vec{r}_j}
\label{eq:PP}\hat{P}(\vec{K}\rightarrow\vec{K}+\vec{G})\Psi_{\vec{K}}(\vec{r})
\end{eqnarray}
The projected function has the appropriate PBZ Bloch wavevector.  The proof looks a little more trivial than it is.  One has to be careful
that it is  legitimate to relabel the ${\cal N}$ PC translations $\vec{r}_i+\vec{r}_j$ as 
PC translations $\vec{r}_k$, because often $\vec{r}_k$
may lie outside the SC.  Is it legitimate just to subtract the appropriate SC translation $\vec{R}_k$ to make $\vec{r}_k$ lie in the
interior of the SC?  The answer is yes, for a subtle reason.  The same difficulty occurs in proving $\hat{P}^2=\hat{P}$, or
\begin{equation}
\hat{P}(\vec{K}\rightarrow\vec{K}+\vec{G})^2=\frac{1}{{\cal N}^2}\sum_{i,j} \hat{T}(\vec{r}_i+\vec{r}_j)
e^{-i(\vec{K}+\vec{G}) \cdot (\vec{r}_i+\vec{r}_j)}
\label{eq:P2}
\end{equation}
The simple re-labeling, that allows the double sum in Eq.(\ref{eq:P2})
to be ${\cal N}$ times the single sum in Eq.(\ref{eq:P}), has to be considered with skepticism;
it is not true in general, but only true because the $\hat{P}$ operator can operate only on functions with SC Bloch
wavevector $\vec{K}$.  When doing so, any part where $\vec{r}_i+\vec{r}_j$ lies outside the SC has the following effect.
This translation is rewritten as $\vec{r}_k +\vec{R}_k$, where $\vec{R}_k$ is a SC translation.  The corresponding translation
operator becomes $\hat{T}(\vec{r}_k)\hat{T}(\vec{R}_k)$.  The translation by $\vec{R}_k$ generates the Bloch phase
$\exp(i\vec{K}\cdot\vec{R}_k)$.  This cancels a corresponding phase $\exp(-i\vec{K}\cdot\vec{R}_k)$ in Eq.(\ref{eq:PP}),
and, together with the identity $\exp(i\vec{G}\cdot\vec{R}_k)=1$, proves the required theorems.

The final step is to show that the sum of all ${\cal N}$ projectors $\hat{P}(\vec{K}\rightarrow\vec{K}+\vec{G})$ is the unit operator, or
equivalently, the translation $\hat{T}(0)$ by zero.  This follows easily from Eq.(\ref{eq:Sum1}).  Therefore, any function with any
SC Bloch symmetry $\vec{K}$ has a unique (representation-independent) 
decomposition into ${\cal N}$ functions of the higher PC translational symmetries
$\vec{K}+\vec{G}_i$.  This actually does not require having an approximate higher translational symmetry.  The translational cells
can be arbitrarily divided into subcells that replicate the full crystal by additional translations, and Bloch functions can be 
separated into components, each of which has some higher translational symmetry.

\subsection{Unfolding formula}
\label{sec:formula}

The previous section suggests a formula
\begin{equation}
W_{\vec{K}J}(\vec{G})=\int d^3\vec{x}|\hat{P}(\vec{K}\rightarrow\vec{K}+\vec{G})\Psi_{\vec{K}J}(\vec{x})|^2,
\label{eq:weights}
\end{equation}
{\it i.e.}, that the PBZ $\vec{K}+\vec{G}$-weight is the norm of the $\vec{K}+\vec{G}$-projected part of the 
SC wavefunction, $\Psi_{\vec{K}J}$.  Using $\hat{P}^2 = \hat{P}$, this is equivalent to
\begin{eqnarray}
W_{\vec{K}J}(\vec{G})&=&\langle\vec{K}J|\hat{P}(\vec{K}\rightarrow\vec{K}+\vec{G})|\vec{K}J\rangle \nonumber \\
&=& \frac{1}{{\cal N}}\sum_{j=1}^{\cal N} \langle \vec{K}J|\hat{T}(\vec{r}_j)|\vec{K}J\rangle 
e^{-i(\vec{K}+\vec{G}_i ) \cdot\vec{r}_j}.
\label{eq:weight}
\end{eqnarray}
The sum rule 
\begin{equation}
\sum_i W_{\vec{K}J}(\vec{G}_i) = 1 
\label{eq:Sum3}
\end{equation}
follows from Eq.(\ref{eq:Sum1}).  This could be used on numerical supercell wavefunctions $\Psi_{\vec{K}j}$
to indicate how similar is a numerical $\vec{K}$-state to a $\vec{k}=\vec{K}+\vec{G}_i$ 
Bloch state of the underlying PC.    The formula uses only the eigenfunctions of the SC, with no reference states needed.

\subsection{relation to spectral function}
\label{sec:spectral}

One of the motivations of KBL \cite{Ku} for unfolding was to facilitate
comparison with  angle-resolved photoemission spectra.  These provide an approximate 
measure of the ``spectral function," $A(\vec{k},E)$, rigorously 
definable as  



 %
 \begin{equation}
A(\vec{k},E)=(-1/\pi) \tr {\rm Im} \hat{G}(\vec{k},E+i\delta).
\label{eq:spect}
\end{equation}
 The Green's function matrix $\hat{G}_{nn^\prime}(\vec{k},E)$  is the time-to-frequency transform
of $-i\langle \hat{T} c^\dagger_{\vec{k}n} (t) c^{}_{\vec{k} n^\prime}(0)\rangle$. 
 Any complete set of Bloch functions can be used to define $\hat{G}$ and give a spectral function $A$. 
If the set is orthonormal, invariance of the trace under unitary transfomations shows
that the spectral function is invariant.  
Computations for a disordered alloy can be done using a ``supercell,''
an artificial periodic construct that enables computation.  A sharp $\vec{K}$-space spectrum 
is computed for representative configurations.  The sharpness is an artifact, 
to be destroyed by  averaging over an ensemble of supercells occupied
by different representative local alloy configurations.  The result, after unfolding and averaging, is a
$\vec{k}$-space description.  Single-particle Bloch-like states have energies $\epsilon_{\vec{K}J}$
and unfolding weights $W_{\vec{K}J}(\vec{G})$.  These define 
average values $\epsilon(\vec{k})$ where $\vec{k}=\vec{K}+\vec{G}$.  The distribution of weighted unfolded
states determines a spectral function $A(\vec{k},E)$.
Green's function language allows us to interpret this {\it via} a complex self-energy $\Sigma(\vec{k},E)$.
Green's function theory tends to motivate perturbative understanding of $\Sigma(\vec{k},E)$.
Non-perturbative computational construction has only recently  been implemented. \cite{Ku,Haverkort} 
Computations, in order to sample realistically the multiple complicated alloy configurations, 
need large supercells.   A single, very large SC computation may approach
the ``self-averaging'' limit,  and yield a good approximation
to $A(\vec{k},E)$ by unfolding.

Consider a particular supercell with known Bloch eigenstates $|\vec{K}J\rangle$.  Let there be a primitive cell with 
a complete Bloch basis $|\vec{k}n\rangle$.  Note that the states $|\vec{k}n\rangle$ are not
required to be eigenstates of anything except translations, but they are required to be complete in the
space of translational quantum number $\vec{k}$.  
To any $\vec{k}$ in the PBZ belongs a unique $\vec{K}$ in the SBZ.
(However, to each SBZ $\vec{K}$ there are ${\cal N}$ PBZ $\vec{k}$'s.)
Assuming the eigenstates $|\vec{K}J\rangle$ also to be complete (in the $\vec{K}$-subspace,
which includes Bloch functions of $\vec{k}=\vec{K}+\vec{G}$-symmetry, for any $\vec{G}$), 
we can write expansions
\begin{equation}
|\vec{k}n\rangle=\sum_J |\vec{K}J\rangle \langle\vec{K}J|\vec{k}n\rangle,
\label{eq:expk}
\end{equation}
\begin{equation}
c_{\vec{k}n}=\sum_J  \langle\vec{K}J|\vec{k}n\rangle c_{\vec{K}J},
\label{eq:crea}
\end{equation}
where $c_{\vec{k}n}$ is a destruction operator for the PBZ Bloch basis function, and similarly
$c_{\vec{K}J}$ for the SBZ Bloch eigenstate.  The single-particle approximation is assumed,
and no ensemble averaging is yet performed on the SC states.  Therefore the Green's
function of the SC is
\begin{equation}
\hat{G}_{JJ^\prime}(\vec{K},E)=(E-\epsilon_{\vec{K}J})^{-1} \delta_{JJ^{\prime}}
\label{eq:GKJ}
\end{equation}
An alternate Green's function is defined using the PBZ Bloch basis,
\begin{eqnarray}
\hat{G}_{nn^\prime}(\vec{k},E)&=& -i\int_0^\infty dt e^{-iEt/\hbar}
\langle \hat{T} c_{\vec{k}n^\prime} (t)  c_{\vec{k}n}^\dagger (0) \rangle
\nonumber \\
&=&\sum_J \langle \vec{k}n^\prime | \vec{K}J\rangle 
\langle \vec{K}J| \vec{k}n\rangle (E-\epsilon_{\vec{K}J})^{-1}.
\label{eq:Gkn}
\end{eqnarray}
This alternate Green's function contains all the information of Eq.(\ref{eq:GKJ}),
except unfolded into the larger PBZ, with the full details of the exact SC eigenstates
obscured in the matrix nature and in the complex coefficients  given by overlap matrices.
Now we can use Eq.(\ref{eq:spect}) to define a spectral function and corresponding weight,
\begin{equation}
A(\vec{k},E)= \sum_J W_{\vec{K}J}(\vec{G}) \delta(E-\epsilon_{\vec{K}J}),
\label{eq:AW}
\end{equation}
\begin{equation}
W_{\vec{K}J}(\vec{G})= \sum_n |\langle\vec{K}J|\vec{k}n\rangle|^2 .
\label{eq:W}
\end{equation}
The weight $W_{\vec{K}J}(\vec{G})$ corresponds to the KBL definition,
up to their assumption that the Wannier functions have the translational symmetry
of the primitive cell unit.
If the Bloch basis $|\vec{k}n\rangle$ is complete, the
weight of Eq.(\ref{eq:W}) is identical to $W_{\vec{K}J}(\vec{G})$ of Eq.(\ref{eq:weight}).
The proof follows (for complete, orthonormal basis sets) 
by inserting  $\hat{1}_{\vec{K}}=\sum_{nm}|\vec{k}+\vec{G}_m,n\rangle
\langle\vec{k}+\vec{G}_m,n|$.  The completeness relation in the SC $\vec{K}$-subspace
requires summing over the ${\cal N}$ PBZ vectors $\vec{k}+\vec{G}$ that map into $\vec{K}$.
This is inserted on either side of the translation operator in Eq.(\ref{eq:weight}).  
Then simple theorems, including Eq.(\ref{eq:Sum2}),  prove the equivalence
to Eq.(\ref{eq:W}).

\section{Notes and possible applications}
\label{sec:notes}

An unfolded spectrum does not contain the full information that was 
in the SC spectrum.  The full PC Green's function $\hat{G}_{JJ^\prime}(\vec{K})$ contains
full information, but its trace, used to find the spectral function and the unfolding
weights, does not.  The purpose of unfolding is to simplify in a way that provides
physical insight.
The unfolding weight defined in Eq.(\ref{eq:weight}) has some possible advantages.
By avoiding the need for a basis of reference functions in a PBZ, it encourages
more general applications.  For example, phonon spectra can be similarly unfolded.

Among the applications that might be imagined are curious ones of uncertain value
that could be tried.  For example, the spectrum of a crystal with rocksalt structure
could be unfolded from the primitive rocksalt two-atom cell into a smaller simple
cubic one-atom cell.  A primitive fcc crystal could be regarded as a supercell of 
a simple cubic structure which had alternate cells empty, and could be
unfolded into this hypothetical primitive cell.  Diamond structure
could be considered a supercell which has alternating cells with two and zero atoms,
and could thus also be unfolded to a simple cubic cell.

\section{Example: Phonons in 1 D}
\label{sec:phonons}

\begin{figure}[tbp]
\centering
\includegraphics[width=8.5cm]{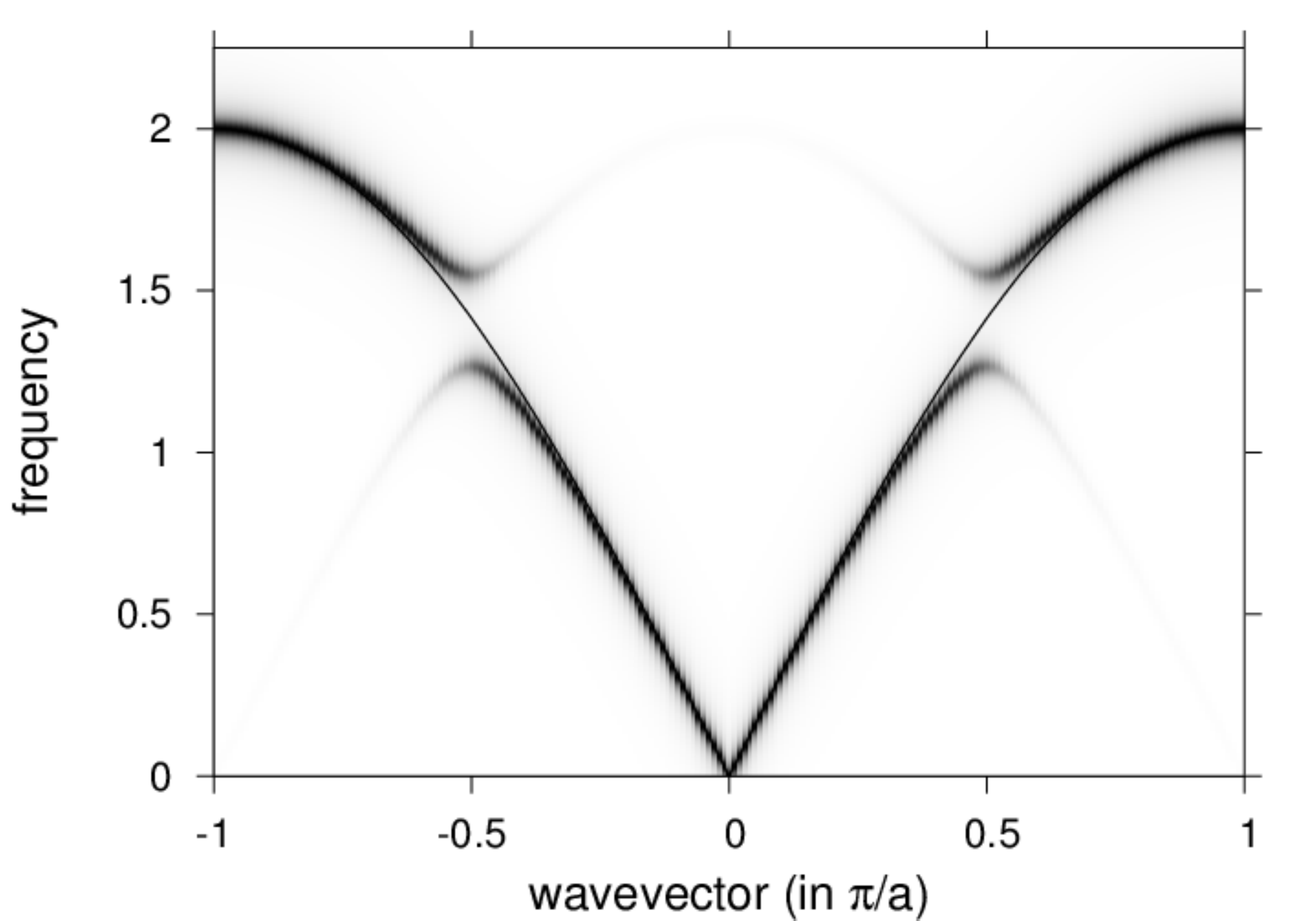}
\caption{Vibrational eigenfrequencies of a diatomic chain, unfolded onto the monatomic Brillouin zone.  The spectrum
of the monatomic chain is also shown as a thin continuous line.  The width of the diatomic dispersion curve 
indicates the amount of weight at the unfolded wavevector. }
\label{fig:ph}
\end{figure}

As a test of the unfolding formula, consider the lattice normal modes of a diatomic chain in one dimension.
The Hamiltonian is
\begin{eqnarray}
H &=& T + V; \ \ \ 
T = \sum_n \left[ \frac{P_{n,1}^2}{2M_1}+\frac{P_{n,2}^2}{2M_2}\right];  \nonumber \\
V &=& \frac{F}{2}\sum_n \left[ (u_{n,1}-u_{n,2})^2 + (u_{n-1,2}-u_{n,1})^2 \right]
\label{eq:H}
\end{eqnarray}
``Atoms'' are alternately of type 1 and 2; they are spaced evenly and interact equally with neighbors on both sides.
The separation is $a$ and the lattice constant is $2a$.  The normal mode spectrum has
two bands (acoustic, labeled ``$-$'', and optic, labeled ``+'') in the BZ $-\pi/2a <K< \pi/2a$. 
Let us regard the two-atom cell of size $2a$ as the supercell (SC), and a one-atom cell of size $a$ as the PC.
The PBZ has $-\pi/a <k< \pi/a$.  In the PBZ there are two vectors, $K$ and $K+G$ (where $G=\pi/a$), for each SBZ
vector $K$.   Eq.(\ref{eq:weight}) gives a prescription for unfolding.   The results are shown in Fig.\ref{fig:ph}.

The algebra is based on textbook procedures \cite{Kittel}, but is slightly tedious.  Details are in
appendix \ref{sec:AppA}.  Bloch's theorem for the SC gives a 2 x 2 dynamical matrix
$\omega_K^2 |s\rangle =\hat{D} |s\rangle$, which has the form
\begin{equation}
\hat{{\sf D}}=\Omega_0^2 \  \hat{1}+
\Omega_1^2 \left( \begin{array}{cc} \cos\theta & -\sin\theta e^{-iKa} \\ -\sin\theta e^{+iKa} & -\cos\theta \end{array} \right).
\label{eq:W2}
\end{equation}
Here $\Omega_0^2$,  $\Omega_1^2$ and $\theta$ are defined by
\begin{equation}
\Omega_0^2 = F  \left(\frac{1}{M_1}+\frac{1}{M_2} \right),
\label{eq:Om0}
\end{equation}
\begin{equation}
\Omega_1^2 = F \left[ \left(\frac{1}{M_1}-\frac{1}{M_2} \right)^2 + \frac{4 \cos^2(Ka)}{M_1 M_2} \right] ^{1/2},
\label{eq:Om1}
\end{equation}
\begin{equation}
\sin\theta = \frac{2F\cos(Ka)/\sqrt{M_1 M_2}}{\Omega_1^2},
\label{eq:th1}
\end{equation}
The eigenfrequencies are $\omega_{\pm}^2=\Omega_0^2 \pm \Omega_1^2$, and the 
weights turn out to be
\begin{eqnarray}
W_{K\pm}(0) &=& \frac{1}{2} [1\mp \sin\theta ] \nonumber \\
W_{K\pm}(\pi/a) &=& \frac{1}{2} [1\pm \sin\theta ].
\label{eq:wph}
\end{eqnarray}
The two branches $K+ $ and $K-$ unfold to the wavevectors $k=K$ and $k=K+\pi/a$.
Over most of the spectrum, a normal mode belongs mostly to one or the other $k$-point,
except near the SC BZ boundary $K=\pi/2a$, where $\sin\theta\rightarrow 0$ and modes belong
equally to both PBZ $k$-vectors, as shown in Fig.(\ref{fig:ph}).  
If the masses evolve to equality ($M_1=M_2$), the weights properly
unfold the spectrum into the single band $\omega_k=(2F/M)|\sin(ka/2)|$.  It does this because the weights $W_{K,-}$ are 1
in the first SBZ and zero in the second, and opposite for $W_{K,+}$.

\section{Slab states unfolded to Bloch states} 
\label{sec:slabs}
 
\begin{figure}[tbp]
\centering
\includegraphics[width=8.5cm,angle=270]{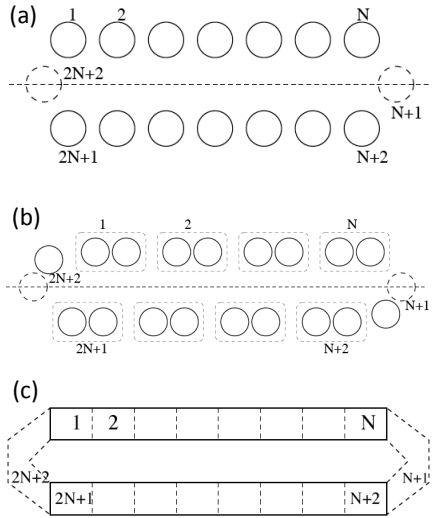}
\caption{Slabs extended cyclically to facilitate unfolding.  (a) is the simple slab of Eq.\ref{eq:1dnn}.  (b) is the diatomic slab with 
a single dangling atom on one end and a corresponding Shockley state.  (c) is the generic slab.}
\label{fig:slabs}
\end{figure}

DFT  calculations are often done for ``slabs,'' cells with a small period in $x$ and $y$ directions, 
and a long period in the $z$ direction.  Examples include modeling stacking faults or artificial multilayers.  A 
common application is for surfaces, where the slab contains significant vaccuum space, 
with two solid/vaccuum interfaces (surfaces.)
Electron chemistry causes surface atoms to relax to new ground state positions.  The surface alters the electronic structure.
Most of the resulting electronic eigenstates are much like the eigenstates of a bulk crystal, except modified near
the surface.  A few states may be localized at surfaces.  For example, dangling bonds are likely to be built from localized surface states. 
Unfolding can be used to illuminate the character of the slab states.
The presence of a vacuum layer creates a new situation.  Is there a natural definition of unfolding in such a case? 
To answer this, it is appropriate to look at specific examples.  The two examples given below lead to a
particular way of dealing with the problem.  One wants to exhibit the standing wave nature of the states in the 
slab's interior (referred to as ``bulk'' states.)  Since we care about electron states that are bound inside the slab,
the standing waves are analogous to classical waves on a string with ``hard wall'' boundary conditions.  That is,
they have the character of sine-wave states $|\vec{k}n\rangle_s=(|\vec{k}n\rangle - |-\vec{k}n\rangle)/\sqrt(2)$, rather than cosine
waves, $|\vec{k}n\rangle_c$, as occur with open-end boundary conditions.  There are only half as many such states as there
are $\vec{k}$-vectors in the BZ, consistent with restricting  $\vec{k} = (\vec{k}_{\parallel},k_z)$-vectors 
to $k_z >0$.  This reduces the number of states by 2.  But we need to describe ${\cal N}$ states.  Hard wall eigenstates
are spaced twice more densely in $k_z$ than those selected by periodic boundary conditions.  
This observation is one guide to how unfolding must be done.

\subsection{Monatomic linear chain}

The first example is a finite one-dimensional
chain of ${\cal N}$ atoms, treated in orthogonal tight-binding approximation, with a single s-orbital per atom.
The s-orbital on the $\ell^{\rm th}$ atom is denoted $|\ell\rangle$.
The single-electron states are  $|\psi\rangle=\sum_{\ell} c_{\ell}|\ell\rangle$, and the eigenstates obey $\hat{H} | \psi \rangle
=E | \psi \rangle$, or
\begin{equation} 
-t \left( \begin{array}{ccccccc} 0 & \gamma & 0 & \hdots & 0 & 0 & 0 \\ \gamma & 0 & 1 & \hdots & 0 & 0 & 0 \\ 
0 & 1 & 0 & \hdots & 0 & 0 & 0 \\ \vdots & & & & & & \vdots\\
0 & 0 & 0 & \hdots & 0 & 1 & 0 \\
0 & 0 & 0 & \hdots & 1& 0 & \gamma \\  0 & 0 & 0 & \hdots & 0 & \gamma & 0 \end{array} \right)
\left( \begin{array}{c} c_1 \\ c_2 \\ c_3 \\ \vdots \\ c_{{\cal N}-2} \\ c_{{\cal N}-1} \\ c_{{\cal N}} \end{array} \right) = E
\left( \begin{array}{c} c_1 \\ c_2 \\ c_3 \\ \vdots \\ c_{{\cal N}-2} \\ c_{{\cal N}-1} \\ c_{{\cal N}} \end{array} \right) 
\label{eq:1dnn}
\end{equation}
The hopping matrix element to nearest neighbor is $-t$; farther neighbor hopping is neglected.  
At the surface, the matrix element is enhanced by a factor $\gamma$.  
First, let us solve this for the special case $\gamma=1$.  This can be done by noticing that it
``inherits'' standing-wave states from the periodic solutions of the cyclic chain with $2{\cal N}+2$ atoms.  
Those are the usual traveling wave Bloch eigenstates, 
\begin{equation} 
|k\rangle = \frac{1}{\sqrt{2{\cal N}+2}}\sum_{\ell} e^{ik\ell} |\ell\rangle
\label{eq:2NBloch}
\end{equation}
with eigenenergy $E(k)=-2t\cos(k)$ and $k$ quantized as $2\pi m/(2{\cal N}+2)$.  The $2{\cal N}+2$ distinct integers $m$
lie in the range $-{\cal N}\le k\le {\cal N}+1$.  Consider now the special standing-wave states
\begin{equation} 
|k\rangle_s = \frac{1}{\sqrt{2{\cal N}+2}}\sum_{\ell=1}^{2{\cal N}+2} \frac{e^{ik\ell}-e^{-ik\ell}}{\sqrt{2}i }|\ell\rangle.
\label{eq:2NBloch}
\end{equation}
This is both a standing sine-wave eigenstate  of the cyclic chain (for $\gamma=1$), plus, 
because of $k$-quantization, it vanishes
on the special atoms $\ell = 2{\cal N}+2 \equiv 0$ and $\ell = {\cal N}+1$.  This means that all such wave-functions
have two distinct parts separated by nodes, and that each separate part is an eigenfunction of the finite
chain of ${\cal N}$ atoms with $\gamma=1$, after renormalizing by $\sqrt{2}$.  That is, the finite ${\cal N}$-atom chain
with $\gamma=1$ has a complete orthonormal set of ${\cal N}$ standing-wave eigenstates,
\begin{equation}
c_{\ell}(k)=\langle \ell | k \{{\cal N},\gamma=1\}\rangle = \frac{\sqrt{2}}{\sqrt{{\cal N}+1}}\sin{k\ell}.
\label{eq:sinwave}
\end{equation}
where $k$ is quantized as $k=\pi m/({\cal N}+1)$, and where the integer $m$ is non-negative, $m=1,2,\ldots,{\cal N}$.
This choice satisfies Eq.(\ref{eq:1dnn}).
The energies of these states are $E = E(k)=-2t\cos(k)$, exactly the same as for the ``bulk'' band structure.
The wavevectors with $m=0$ or ${\cal N}+1$ (corresponding to $k=0$ or $\pi$) do not work because 
the corresponding state vanishes on every atom.
There is a close analog to vibrational standing waves with hard-wall
boundary conditions.

Now consider what happens when $\gamma$ is allowed to vary.  The simplest case is if $\gamma$
is set to zero.  The two end atoms are now decoupled, and the remaining coupled chain
has ${\cal N}-2$ atoms.  The orbitals $|\ell>$ for $\ell=1$ and $\ell={\cal N}$ are now eigenstates
of energy 0, while the ``bulk'' states look just as before, except there are fewer $k$'s, being
quantized in new units $\pi m/({\cal N}-1)$, with $m = 1, 2, \ldots, {\cal N}-2$.  The atoms are re-numbered
so that $2\rightarrow 1$, {\it etc.}, and the quantized $k$'s  are re-defined so that $c_{\ell}$ 
vanishes at $\ell = 0$ and $\ell = {\cal N}-1$.  

\begin{figure}[tbp]
\centering
\includegraphics[width=8.5cm]{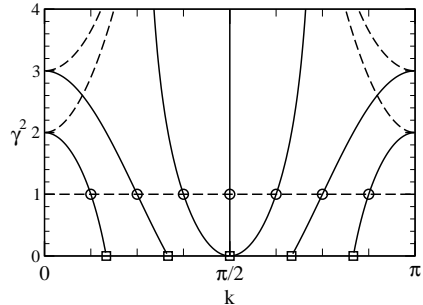}
\caption{$k$-vectors (horizontal; between 0 and $\pi$) evolving as the surface
hopping enhancement factor $\gamma$ (squared and plotted vertically) varies, for a
7-atom single s-orbital tight-binding chain.  Simple periodicity of $k$ is seen in the 
two cases $\gamma=1$ and $\gamma=0$ where surface enhancement is absent and
the simplest bulk behavior is reproduced by slab standing waves.  As the critical value
$\gamma=\sqrt2$ is exceeded, two bulk standing waves disappear, and surface states
appear with energies above and below the bulk bands, symmetric between the two surfaces.  Above a larger
critical value ($\gamma=\sqrt3$ for the 7-atom slab) two more bulk standing waves disappear,
and anti-symmetric surface states appear.  The dashed lines should be imagined moving
perpendicular to the page in the imaginary $k$ direction.  These give the decay rates
$\exp(-\kappa \ell)$ and $\exp(-\kappa({\cal N}+1-\ell))$ of the surface eigenstates.}
\label{fig:kvec}
\end{figure}

For all other values of $\gamma$, besides 0 and 1, careful inspection shows that there continue to be
standing wave eigenstates.  However, $k$-quantization is no longer \cite{Davidson} in simple units of $\pi$
divided by something like ${\cal N}$.  The actual evolution of the standing wave $k$'s, for a 7-atom
chain, is shown in Fig. \ref{fig:kvec}.  There is a critical value, $\gamma_c=\sqrt2$, beyond which
pairs of bulk standing-wave solutions disappear, and new surface states emerge.  It is
simplest to describe if the origin is taken in the middle of the chain.  Then the standing waves
are either even (cosine waves) or odd (sine waves.)  Similarly, the surface states are either
even (cosh) or odd (sinh) combinations of modes that decay exponentially from both surfaces.  Surface states
appear both below the bulk band ($E<-2t$) and above ($E>2t$).  The ones below have
complex $k$-vector $k=i\kappa$ and those above have $k=\pi-i\kappa$.  The value of
$\kappa$ is related by a transcendental equation to the value of $\gamma.$  Their energies
are $\pm 2t\cosh\kappa$.  The symmetric
(cosh) solutions appear at $\gamma^2 >2$ while the odd (sinh) solutions appear at 
slightly larger values $\gamma^2 >2/(1-(2/({\cal N}-1)))$. 

There is a lesson from this that applies to unfolding, namely, 
that in slab calculations, good standing-wave bulk-like solutions continue to exist.
However, they are perturbed by the surface in three ways.  First, k-quantization is twice as dense
as with periodic boundary conditions.  Instead of having ${\cal N}$ solutions between $k = -\pi$ and
$k=\pi$, there are ${\cal N}$, or ${\cal N}-2$, or ${\cal N}-4$ solutions between $k=0$ and $k=\pi$.  Second,
$k$-quantization is no longer in easily predictable multiples of $2\pi/{\cal N}$.  Third, as can be
verified by direct computation, the amplitude of
both bulk and surface states, on
surface atoms, is reduced by $1/\gamma$, when the surface hopping is enhanced by $\gamma$.
Similarly, in real crystals, the surface layers will displace, couplings will be altered, and bulk-like
states will be altered, probably reduced in amplitude, near the surface.

\subsection{Diatomic linear chain}

There is another example, the diatomic chain of ${\cal N}$ molecules, that provides further illumination.  Atoms are paired.  
They have identical on-site energies (set to zero), weaker first-neighbor coupling $-t$ to adjacent molecules, 
and stronger intramolecular $-t^{\prime}=-\gamma t$ coupling within the molecule.  The cyclic chain of $2{\cal N}+2$
molecules is illustrated in Fig.\ref{fig:slabs} (b).  Bloch's theorem reduces this to a 2 x 2 problem.
Denoting the wavefunction coefficients as $u_\ell = u_k \exp(ik\ell)/\sqrt{{\cal N}+1}$ for the left atom of a molecule,
and similarly with $u\rightarrow v$ for the right atom, the Hamiltonian and eigenstates become
\begin{equation}
H_k = - \left( \begin{array}{cc} 0 & |E_k|e^{i\phi_k} \\ |E_k|e^{-i\phi_k} & 0 \end{array}\right),
\label{eq:H2}
\end{equation}
\begin{equation}
\left( \begin{array} {c} u_k \\ v_k \end{array} \right) = \frac{1}{\sqrt{2}}
\left( \begin{array} {c} e^{i\phi_k /2} \\ e^{-i\phi_k /2} \end{array} \right), 
\label{eq:ev2}
\end{equation}
%
%
with $k$ quantized as $k=2\pi m/(2{\cal N}+2)$ and  energy eigenvalues $E_k = \pm |E_k|$,
with $|E_k|=|t|(a_k^2 + b_k^2)^{1/2}$ and $\tan\phi_k = a_k/b_k$.  The parameters
are $a_k = \gamma\sin k$ and $b_k=1 + \gamma\cos k$.
There are two bands symmetrically
distributed around $E=0$.  Once again, standing sine-wave eigenstates can be made, and
can be chosen to vanish on two oppositely placed atoms, as shown in Fig.\ref{fig:slabs}(b).
The sine-waves are eigenstates for the separated half-chains with the nodal atoms eliminated.
Next to each nodal atom is an unpaired atom that remains.  Each separate chain thus has 
${\cal N}$ molecules, plus a single weakly-coupled adatom on one side of the chain.  
The $2{\cal N}+1 \times 2{\cal N}+1$ Hamiltonian matrix for the upper separated chain is

\begin{equation} 
-t \left( \begin{array}{cccccccc} 0 & 1 & 0 & 0 & \hdots & 0 & 0 & 0 \\ 1 & 0 & \gamma & 0 & \hdots & 0 & 0 & 0 \\ 
0 & \gamma & 0 & 1 & \hdots & 0 & 0 & 0 \\ 0 & 0 & 1 & 0 & \hdots & 0 & 0 & 0 \\
\vdots & & & & & & & \vdots\\ 
0 & 0 & 0 & 0 & \hdots & 0 & 1 & 0 \\
0 & 0 & 0 & 0 & \hdots & 1& 0 & \gamma \\  0 & 0 & 0 & 0 & \hdots & 0 & \gamma & 0 \end{array} \right)
\left( \begin{array}{c} v_0 \\ u_1 \\ v_1 \\ u_2 \\ \vdots \\ v_{{\cal N}-1} \\ u_{{\cal N}} \\ v_{{\cal N}} \end{array} \right) = E
\left( \begin{array}{c} v_0 \\ u_1 \\ v_1 \\ u_2 \\ \vdots \\ v_{{\cal N}-1} \\ u_{{\cal N}} \\ v_{{\cal N}} \end{array} \right)
\label{eq:diann}
\end{equation}
There are ${\cal N}$ positive $k$'s with $2{\cal N}$ corresponding standing sine-waves (the 2 being
from the two orbitals per molecule)
of the $2{\cal N}+2$-molecule cyclic chain.  These generate states
for the ${\cal N}+1/2$-molecule finite chain.
Specifically, there are ${\cal N}$ states with energy $-|E_k|$ and ${\cal N}$ with energy $+|E_k|$, two for each intact
molecule of the upper separated chain.  But there are $2{\cal N}+1$ atoms on this chain, so one
state remains to be found.  It is a Shockley-type surface state \cite{Davidson,Shockley} lying exactly at mid-gap, $E=0$.
Its eigenfunction, for the upper chain, has coefficients $v_\ell=(-1/\gamma)^\ell$ and 
$u_\ell =0$.  That is, it is zero on the right atom of each molecule, and decays with
complex wavevector $k=\pi+i\log\gamma$.  It
alternates in sign and decays as $\exp(-\ell\log\gamma)$, going to the right from the leftmost atom
($\ell=1$).  This remains a
midgap eigenstate for all $\gamma\ge 1$.

\subsection{Slab unfolding}

How therefore should we ``unfold'' such slab calculations?  In the real 3d world of surfaces, slab states 
have $\vec{k}$'s quantized in the $x$ and $y$ directions parallel to the slab, but perpendicular,
only $k_z = 0$ states are of interest.  Any dispersion in $k_z$ derives from undesired, hopefully small,
interaction between the slab and its periodic images.  These images were included only for
computational convenience, to permit the use of periodic codes.  
The true slab states are either surface states, or else they are extended,
and hence, standing waves.  Even though they
have quantum number $k_z$=0 from the band-structure code, nevertheless, the extended states relate to 
superpositions of bulk Bloch states with equal contributions from $k_z$ and $-k_z$ for some value of $k_z$.     
The valence and low-lying conduction standing-wave
states are bound in the potential well of the solid.  Thus they have sine-wave rather than 
cosine-wave character.  We wish to know the approximate relevant $k_z=|k_z|$.
Their weights will be the same at $+k_z$ and $-k_z$.  
The maximum value of $k_z$ is certainly $\pi/c$ where $c$ is the (atomic scale) layer spacing within the slab.  But
the $k_z$ values are definitely not related to multiples of $2\pi/{\cal N}$ where ${\cal N}$ is the number of
layers.  Instead, their average spacing is $2\pi/(2{\cal N}+2)$.  Precise spacing is only obeyed
for very special model situations.  A standing wave of a
thick slab should have a sharply defined, but slightly unpredictable, $k_z$ characterizing its
behavior in the interior, and more complicated
wavefunctions near the boundaries, predictable only by detailed calculation.

%
\begin{figure}[tbp]
\centering
\includegraphics[width=8.5cm]{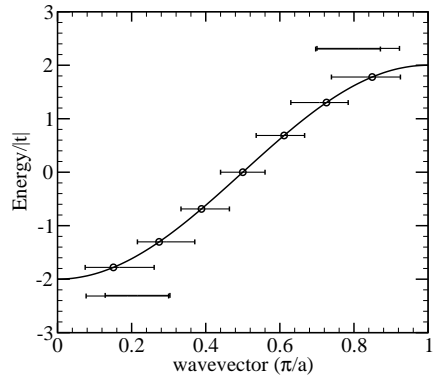}
\caption{Unfolded bands of an 11 atom tight-binding one-band model.  Surface hopping is enhanced
by a factor of $\gamma=2$.  Symmetric and antisymmetric surface states are essentially degenerate.  They
occur at energies ~$0.32t$ above the top and below the bottom of the bulk bands at $\pm 2t$.  The bulk
band dispersion $E(k)=-2t\cos(k)$ is plotted, and agrees with the wavevectors found for bulk states in the 
slab's interior.}
\label{fig:1btbgam2}
\end{figure}

The procedure that seems most general and sensible is illustrated by Fig.\ref{fig:slabs}(c).  The slab inherits an
approximate periodicity $c$ in the perpendicular direction, and has ${\cal N}$ well-defined layers plus possible adatoms
or reconstruction at the surface.  This gives an approximate thickness ${\cal N}c+\delta$.  
Imagine instead a cycle of length $(2{\cal N}+2)c$.
If the surface is thick with adatoms, $(2{\cal N}+4)c$ might be preferable.
Reflect the eigenstates of the actual slab through the separating midline of
Fig. \ref{fig:slabs}.   This generates a candidate continuation onto the $(2{\cal N}+2)$-loop.  But because of the
hard-wall standing wave nature of the bulk states, and sine-like disappearance of the bulk states as they
merge into the vacuum, the reflected wavefunction should have its sign changed to give it sine-like character.
Finally, use the unfolding algorithm
explained above in Sec. \ref{sec:formula}.  The basis $k$-vectors can be chosen to have 
$k_z = \pi m/({\cal N}+1)$ for $m=1,2,\ldots,{\cal N}$.  These will not exactly conform to the actual 
standing-wave eigenvector wavelengths, but provide a convenient basis set for unfolding.
Surface states will unfold to a broad range of $k_z$'s, while bulk-like standing waves,
for thick slabs, will have weights distributed
more narrowly. 

Finally, it is sensible for plotting $E$ versus $k_z$ to use a horizontal bar to indicate the width in $k_z$
that a state unfolds to.  The mean and rms $k$-values can obviously be defined as $\bar{k}=\sum_k W(k) k/\sum_k W(k)$,
and $\delta k^2=\sum_k W(k) (k-\bar{k})^2 /\sum_k W(k)$.

As an example, Fig. \ref{fig:1btbgam2} shows the results for the single-orbital 1d tight-binding chain of
11 atoms thickness.   The surface enhancement factor was chosen as $\gamma=2$, large enough so 
that both even and odd surface states occur, both above and below the bulk bands.  The layer thickness
of 11 is sufficient that very little splitting occurs between even and odd.  The figure shows by circles the ``exact"
values of $k_z$ that characterize the interior.  These are found by numerical solution of transcendental equations
derived from Eq.(\ref{eq:1dnn}).  Their energies agree with the eigen-energy $-2t\cos(k)$.

\section{silicon (111) slab unfolded}

\begin{figure}[tbp]
\centering
\includegraphics[width=8.5cm]{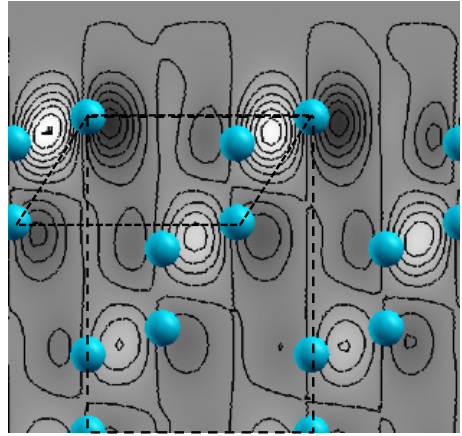}
\caption{Section through the silicon slab, seen along $\vec{a}-\vec{b}$.  The vector $\vec{c} \parallel (111)$ is
vertical, with vacuum at the top.  Vector 
$\vec{a}+\vec{b}$, defined in the text, is horizontal.  Two types of translational cells are shown.  The parallelogram outlines
a diagonal slice through a primitive diamond cell with two silicon atoms.  The rectangle outlines a
slice through a hexagonal cell containing
6 atoms, arranged in three bilayers.  The bilayers stack in the {\it fcc} ABC pattern.
Also shown are contours of the real part of $\psi$ for one of the $p\sigma$ surface bands in the gap, shown by an
arrow in Fig.\ref{fig:si111}.  Contours are evenly spaced; pale and dark grey indicate opposite signs. }
\label{fig:str111}
\end{figure}

We performed an ${\cal N}=18$ double-layer silicon slab calculation using the SIESTA \cite{siesta} implementation
of density-functional theory (DFT), and the PBE-GGA \cite{Perdew} exchange-correlation
potential.   The slab orientation is perpendicular to (111).  
Charge self-consistency was obtained using 128 k-points (all perpendicular to the (111) axis.) 
A cubic lattice constant $a=5.499 \AA$ (compared to experimental $a = 5.43 \AA$)
 minimizes the total energy of the 2-atom bulk primitive cell (PC), using 
 the standard double-zeta-polarized (DZP) basis orbital set with 13 orbitals per silicon atom. The unit 
 cell has 36 atoms, one per layer (two per double layer.)  Translation vectors 
 within the (111) plane are  $\vec{a}=\frac{a}{2} (-1 \,0 \,1)$ and $\vec{b}=\frac{a}{2} (0 \, -1 \, 1)$. 
 Since the aim is to study unfolding, the simplest possible surface is used, namely a completely
 unrelaxed cut between adjacent double layers.  
 The bottom of the slab is symmetrical with the top.  The top-most and  
 bottom-most atoms have one dangling bond each. 
 The shortest translation vector in the (111) direction is $\vec{c}=a(1 \,1 \,1)$.  The vectors ($\vec{a},
 \vec{b}, \vec{c}$) define an hexagonal cell with 6 atoms arranged in three double layers, as shown
 in Fig.\ref{fig:str111}.  The 18 double layers define a section of bulk with vertical distance $6\vec{c}$.
 On top of the slab is a large vacuum layer, and periodic repeats with period $50\vec{c}$.
 
\begin{figure}[tbp]
\centering
\includegraphics[width=8.5cm]{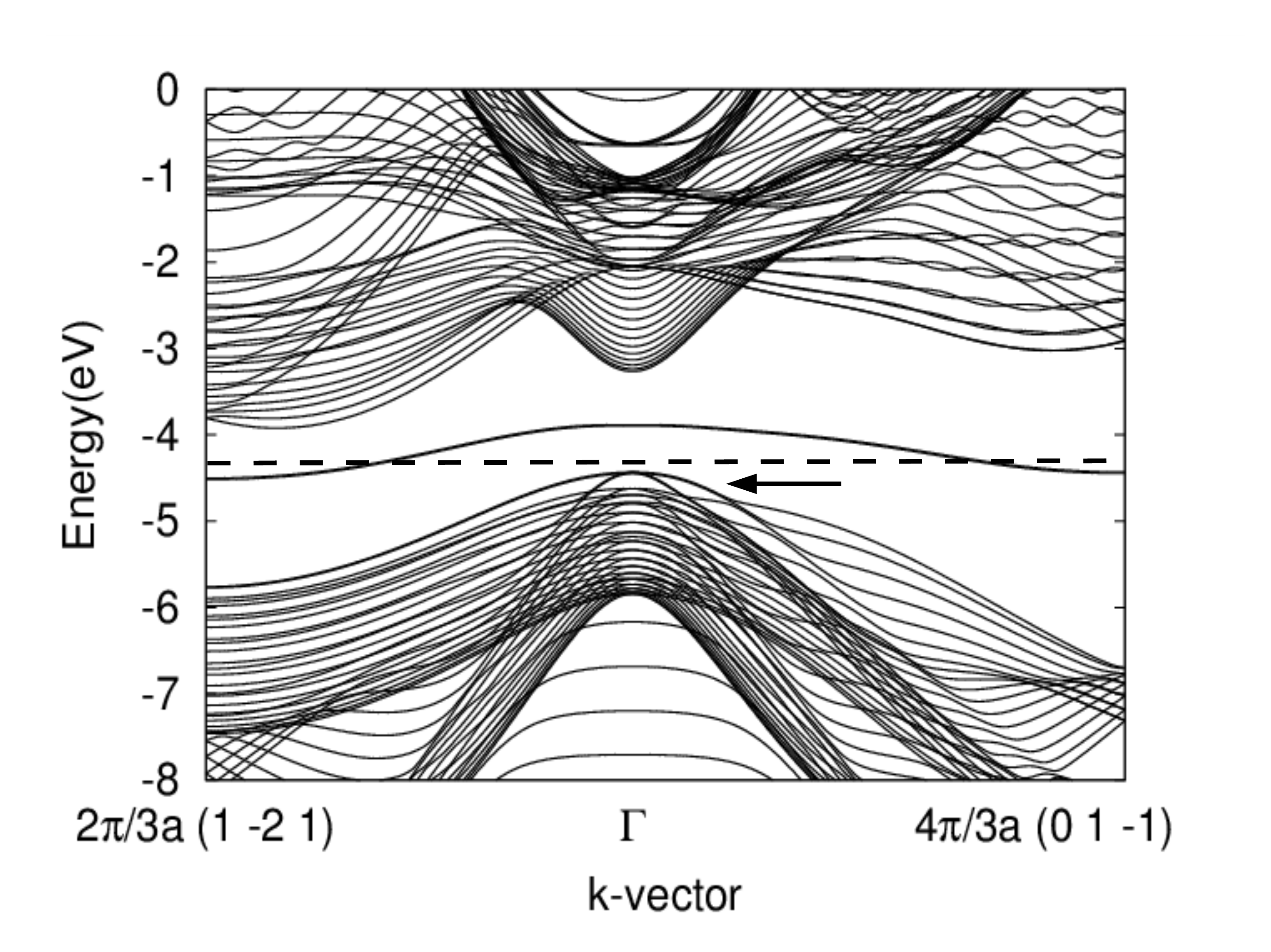}
\caption{Band dispersion for the (111) slab of 18 double layers, with 36 atoms per cell.  Two directions perpendicular to (111) are
shown.  The Fermi level (dashed line) intersects a doubly-degenerate surface state.  This state is a result of one dangling bond per surface
atom (and two surfaces), obtained because atoms are frozen in perfect terminated bulk positions.  The arrow points to occupied surface
states just above the top of the valence band, one of which is shown in contour in Fig.\ref{fig:str111}.}
\label{fig:spectrum}
\end{figure}

 The computed band structure is shown in Fig.\ref{fig:spectrum}.  Note that the Fermi level lies in the middle of
 a doublet of surface states.  The surface is metallic.  It is certainly possible that including spin-polarized options
 in the DFT calculation would have split up and down spins, perhaps creating a magnetic insulator instead of
 a metal.  It did not seem important to test this, because the unrelaxed ideal surface can only be
 regarded as an over-simplified model.

 The primitive unit cell has two atoms.  A conventional description would use vectors $\vec{a}, \vec{b}, \vec{d}$,
 where $\vec{d}=\vec{c}/3-(\vec{a}+\vec{b})/3$ is shown in Fig.\ref{fig:str111} as the slanted side
 of the parallelogram.  It has vertical height $|\vec{c}/3|$. 
 Two comments can be made.  First,
 a 2-atom hexagonal cell defined by $\vec{a}, \vec{b}, \vec{c}/3$ is an allowed unit cell; it translates by 
 multiples of $\vec{a}, \vec{b}, \vec{d}$ to fill space, like bricks, with a horizontal off-set.  Second,
 the primitive cells fill the slab unit cell (supercell) by symmetry translations that slant relative to the $\vec{c}$-axis.
 This forces simple modifications in applying the unfolding algorithm.

SIESTA gives expansion coefficients, denoted here by $\langle N | J \rangle$, 
 for the slab eigenstate with band index J, in terms of the slab local basis orbitals $|N\rangle$. 
 We examine only eigenstates with $\vec{K}=0$, and ask how their approximate translational 
 symmetry unfolds onto wavevectors parallel to $\vec{c}$.   The slab comes from a section of bulk
 of thickness $18c/3$ where $c=|\vec{c}|=\sqrt{3}a$. The longest standing-wave wavelength
is then approximately $\lambda_{\rm max}=2\cdot 19 \cdot (c/3)$.   The factor 2 is because the
longest sine-wave has only a half wave-length in the slab.  The factor 19 is used instead
of 18, following Sec.\ref{sec:slabs}.  The extra distance $c/3$ estimates the additional distance for a typical sine-wave
state to vanish at the edges.  The $k_c$-character of states should be approximately quantized in
units $k=m\cdot(2\pi/\lambda_{\rm max})$, $m$ going from $1$ to ${\cal N}=18$, as required to represent the states
of the 18 double layers.

The unfolding algorithm of Sec.\ref{sec:slabs} and KBL \cite{Ku} is implemented as follows.  
First, a mirror copy
of the slab is positioned adjacent, with the extra separation $c/3$.  Second, wavefunctions $|J\rangle$
must be extended from the original slab to its mirror.  Third, this $2{\cal N}+2$ double-layer slab
is repeated periodically (vacuum is neglected except for the two extra $c/3$ slices.)
The mirror image slab orbital $N^\prime$ corresponding to orbital $N$ of the real slab is assigned 
the expansion coefficient $\langle N^\prime | J \rangle = - \langle N | J \rangle$.
The extra space $c/3$ contains a missing double-layer, whose 2 missing atoms contain contributions
to state $|J\rangle$  with $\langle N | J \rangle=0$.  All coefficients 
are divided by $\sqrt2$ to maintain normalization. 
Fourth, a mapping is constructed between orbitals of the slab and the 26 ``reference orbitals'' that
form the SIESTA basis states of the primitive 2-atom cell. 
The transverse components of the positions are ignored, as only the 
relative position in the (111) direction is relevant for $k_z$ unfolding.

The slab eigenstates are unfolded onto the $2{\cal N}+2$ wavevectors 
$\vec{k}=(0,0,k_z) = (m\pi/({\cal N}+1)a) (1 1 1)$, 
where $m = -{\cal N},  \ldots, -1, 0, 1, \ldots, {\cal N}+1$. 
Only values $m = 1, \ldots, {\cal N}$ are needed, as confirmed below.
The weights are given by 
\begin{equation}
W_J (\vec{k_z}) =  \sum_{n,n^\prime}  \langle k_z n | J \rangle \langle J| k_z n^\prime \rangle \langle k_z n^\prime | k_z n \rangle. 
\label{eq:nonorthW}
\end{equation}
This is the altered version of Eq.(\ref{eq:W}), the difference arising from the non-orthogonal nature of SIESTA basis functions.
The elements $\langle k_z n | J \rangle$ are given by
\begin{equation}
\langle k_z n | J \rangle = \frac{1}{\sqrt{2{\cal N}}} e^{-i \vec{k_z}\cdot{\vec{r_o}(n)}} \sum_N e^{-i\vec{k_z}\cdot \vec{r}(N)} {\delta}_{nn(N)} 
\langle N | J \rangle.
\label{eq:SW}
\end{equation}
The sum over $N$ runs over all orbitals in the slab plus mirror slab construction. $\vec{r}(N)$ and 
${\delta}_{nn(N)}$ comprise the two components of the orbital mapping. $n(N)$ is the PC orbital 
corresponding to the slab orbital $N$ and $\vec{r}(N)$ is the position vector connecting $n(N)$ to $N$. 
$\vec{r_0}(n)$ is the position of the orbital $n(N)$ within the PC. 
In the formula for the weight $W_J (\vec{k_z})$, 
$\langle k_z n^\prime | k_z n \rangle$ is the overlap matrix element between the two Bloch basis 
states associated with the local orbitals $n$ and $n^\prime$. This overlap matrix is extracted from a SIESTA 
calculation for the bulk silicon PC. 

\begin{figure}[tbp]
\centering
\includegraphics[width=8.5cm]{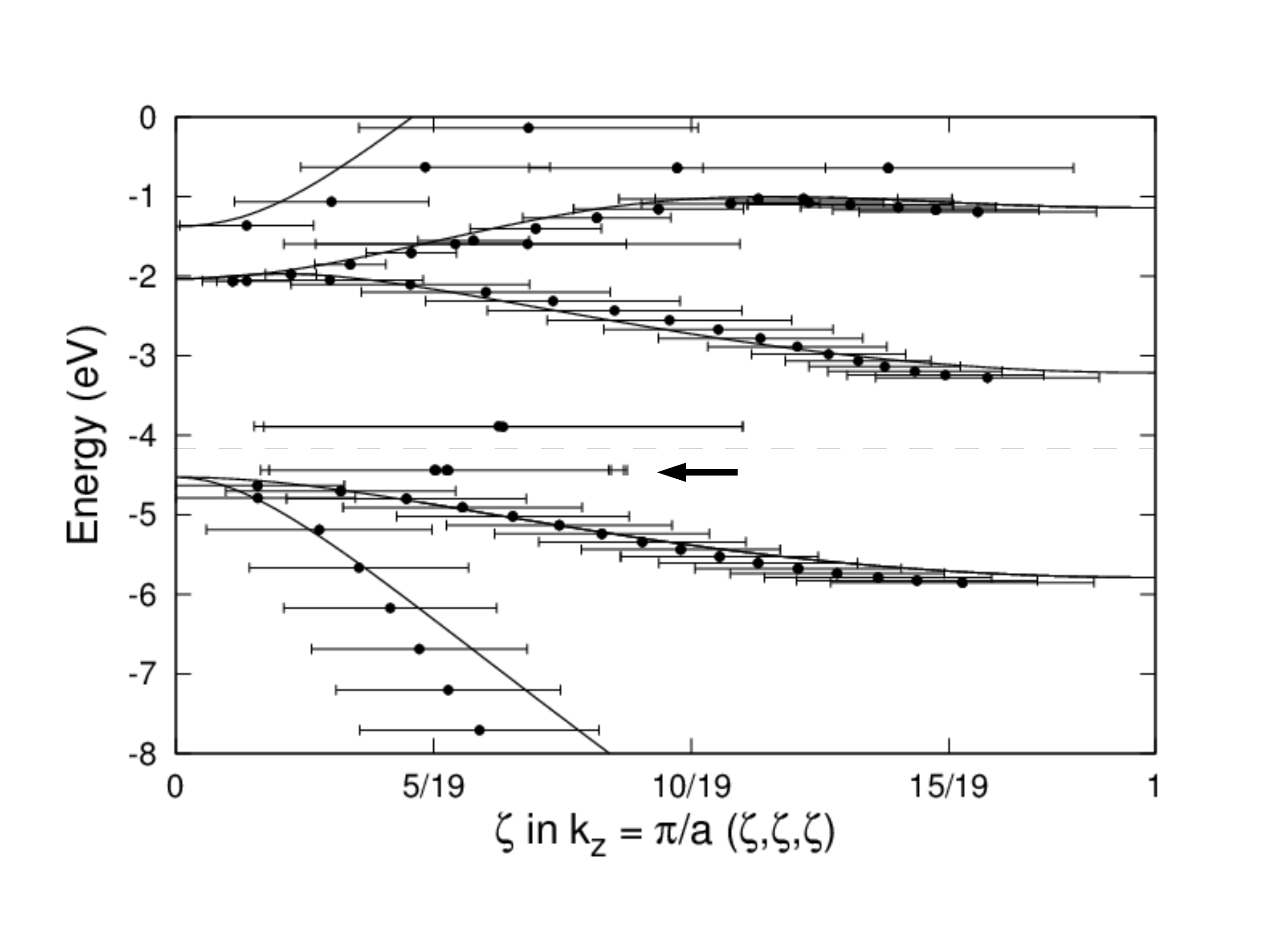}
\caption{The $\vec{k}=(0,0,0)$ bands  of the 18 bi-layer Si (111) slab (the states shown at $\Gamma$ in Fig.\ref{fig:spectrum}) 
are here unfolded to reveal their relation to bulk states with $\vec{k}
=(\pi/a)(\zeta,\zeta,\zeta)$.  Smooth curves show the bulk dispersion in this direction.  Horizontal bars
indicate the position and width of the peak of the distribution in $\zeta$ for each eigenstate of the slab.  The dashed 
line indicates the Fermi energy, and the arrow locates the surface state shown in Fig. \ref{fig:str111}.}
\label{fig:si111}
\end{figure}

The weights for $k_z$ and $-k_z$ are found to be identical, as expected.
The weights for $k_z=0$ and $k_z = \pi$ are zero, so there are 18 independent weights for each 
state $|J\rangle$, as expected.  On average, the sum of 
the unfolding weights $\sum_{k_z} W_J (\vec{k_z})$ satisfies the sum rule given by Eq. (8). 
More specifically, the average value of $\sum_{k_z} W_J (\vec{k_z})  = 0.998 \pm 0.114$. 
The sum rule is not exact, because of non-orthogonality of the SIESTA basis orbitals. 
The plot of E versus $k_z$ in Fig.\ref{fig:si111} shows for each eigenstate near the band gap,
the mean and rms variance of the distribution of $k_z$ that the state unfolds to.

Notice that there are three states visible in the gap in Fig.\ref{fig:si111}, each of which is
actually a doublet.  The position and width of their $k$-distribution reveals that they are paired surface states,
lying on top and bottom surfaces, almost completely de-coupled.  
The higher energy doublet involves (111)-oriented $p$-states, which bond strongly to each other when the
separated surfaces are allowed to recombine.  Upon coupling, one state moves down to the top of the valence band,
while the other moves up, above the anti-bonding $s$ state, in the conduction band.  The lower energy quartet,
when viewed without unfolding, in Fig.\ref{fig:spectrum}, would not be easily recognized as localized, except 
that the unfolded version in Fig.\ref{fig:si111} makes its nature clear.  We examined contour plots of all four
of these states, to confirm the localized nature.  One of these states is shown in Fig.\ref{fig:str111}. 
They are $p$-states oriented parallel to the plane, bonded weakly to neighboring surface atoms.  It is
not obvious that opening the vacuum between bi-layers should so much affect their localization.

Of the higher-lying anti-bonding
states, inside the conduction band, shown  in Fig.\ref{fig:si111},
three states near $-$0.6 eV seem nearly degenerate,
and two of them show anomalous $k_z$-broadening.
These two are actually a quartet of $p$ states, localized (according to contour
plots) on top and bottom surfaces.  They are anti-bonding counterparts to the in-gap surface state quartet discussed above.  
The third state with the smaller $k_z$-distribution is quite different.  It is a single delocalized state of $s$-character,
a member of the rapidly upward-dispersing band of anti-bonding $s$-states. 
The localized quartet  cannot be true surface states, since they do not occur within an energy window
where bulk states with $\vec{k}=(0,0,k_z)$ are missing.  They are degenerate with the rapidly dispersing band,
and are therefore surface resonances.  
Only a larger cell (more than 18 double-layers) could reveal the resonant aspect.
Of course, none of these
appear in actual experimental silicon (111) surfaces.  There is an easy route to a ``$2 \times 1$'' reconstruction \cite{Stroscio},
and a more difficult route to the ground-state $7 \times 7$ reconstruction \cite{Takayanagi}.

\section{Summary and conclusions}

This paper has done two things.  The notion of unfolding has been given a more general interpretation.  A
candidate general definition is made that uses only the states under direct consideration, needing in principle
no basis of Bloch states of the simpler primitive cell.  Only the properties of the complicated states under
partial translation are used.  

The idea of unfolding is then generalized to systems that are finite in one
direction.  Specifically, we consider a slab, which ``remembers'' or ``inherits'' the
intrinsic periodicity $c$ that would have been exact in bulk.
The vacuum part of the slab is ignored, and the occupied part regarded as a piece of a periodic medium
with ${\cal N}+1$ (rather than the nominal ${\cal N}$) vertical repetitions of the bulk structure, the last repetition
being vacant of atoms.  To accomodate
the half-size $k_z$ quantization of standing waves, the slab is then mirrored to a $2{\cal N}+2$-layered slab
with periodic boundary conditions.  Wavefunctions on the mirror image are negatives of their mirror counterparts.
Finally, these modified states are unfolded in the normal way,
revealing the hidden $k_z$ character of the states.  Localized surface states, and partially localized
surface resonances, are revealed.  Not all of these properties were expected or discovered from the 
conventional dispersion plot (Fig.\ref{fig:spectrum}), but were uncovered by the unfolding algorithm.

\appendix
\section{1-d phonons}
\label{sec:AppA}

The Hamiltonian of the diatomic chain of vibrating atoms is Eq.(\ref{eq:H}).
According to Eq.(\ref{eq:weight}), the weights that we want for this system are
\begin{equation}
W_{K,\pm}(B)=\frac{1}{2}\left[1+\langle K,\pm|\hat{T}(a)|K,\pm\rangle e^{-i(K+B)a} \right],
\label{eq:Wph}
\end{equation}
The chain has periodicity $2a$.  The translation $\hat{T}(a)$ gives the ``inherited'' approximate translational symmetry.
The factor $\exp(-iBa)$ is either $+1$ (when $K$ is unfolded onto itself by $B=0$) or $-1$ (when $K$ is unfolded
onto $K+B=K+\pi/a$ in the ``second'' BZ.)

First, make the standard SBZ Bloch-wave (Fourier) transform for displacements, $u_{n,i}\rightarrow u_i(K)$.
Looking for normal modes that oscillate sinusoidally with frequency $\omega_K$, Newton's laws become
\begin{equation}
\omega_K^2{\hat{\sf M}}|u\rangle = {\hat{\sf F}}|u\rangle  \ \ \ |u\rangle = \left( \begin{array}{l} u_1(K) \\ u_2(K) \end{array} \right).
\label{eq:u}
\end{equation}
The mass matrix ${\hat{\sf M}}$ is diagonal with elements $M_{ij}=M_i \delta_{i,j}$.
It is convenient to make the mass-weighting transform $|s\rangle = {\hat{\sf M}}^{1/2}|u\rangle$, which converts Eq.(\ref{eq:u}) to the 
standard Hermitean problem
\begin{equation}
\omega_K^2 |s\rangle ={\hat {D}} |s\rangle,
\label{eq:s}
\end{equation}
where the mass-weighted force matrix $\hat{D}$ is
\begin{equation}
\hat{D}=F
\left( \begin{array}{cc} \frac{2}{M_1} & -\frac{2\cos( Ka)}{\sqrt{M_1 M_2}} e^{-iKa} \\
-\frac{2\cos( Ka)}{\sqrt{M_1 M_2}} e^{+iKa} &  \frac{2}{M_2}  \end{array}\right).
\label{eq:Wph}
\end{equation}
It is convenient to rewrite this as done in Sec. \ref{sec:phonons} Eqs.(\ref{eq:W2}-\ref{eq:th1})
The eigenvectors of the dynamical matrix (Eq.\ref{eq:W2}) are
\begin{equation}
|s-\rangle=\left(\begin{array}{c}\sin(\frac{\theta}{2})e^{-iKa/2} \\ \cos(\frac{\theta}{2})e^{+iKa/2}\end{array}\right) \ \ 
|s+\rangle=\left(\begin{array}{c}\cos(\frac{\theta}{2})e^{-iKa/2} \\ -\sin(\frac{\theta}{2})e^{+iKa/2}\end{array}\right) \ \ 
\label{eq:eigv}
\end{equation}
We now need to translate an eigenvector.  To find the form of $\hat{T}(a)|s_i\rangle$, it is helpful to display the full 
coordinate-space vector $|s\rangle$ of mass-weighted displacements for a Bloch wave with wavevector $K$,
\begin{equation}
|s\rangle = \left(\begin{array}{l}\vdots \\ s_1 e^{2iK(n-1)a} \\ s_2 e^{2iK(n-1)a} \\ s_1 e^{2iKna} \\ s_2 e^{2iKna} \\
s_1 e^{2i(n+1)Ka} \\ s_2 e^{2iK(n+1)a} \\ \vdots \end{array} \right).
\label{eq:fulls}
\end{equation}
This tells us that the rule for translation is
\begin{equation}
\hat{T}(a)\left( \begin{array}{l} s_1(K) \\ s_2(K) \end{array}\right) = \left( \begin{array}{l} s_2(K) \\ s_1(K)e^{+2iKa} \end{array}\right).
\label{eq:Ts}
\end{equation}
The diagonal matrix elements of the translation by $a$ are then
\begin{equation}
\langle s\pm |\hat{T}(a)| s\pm\rangle = \mp \sin\theta e^{+iKa}
\label{eq:ME}
\end{equation}
We can now evaluate the unfolding weights, Eq.(\ref{eq:Wph}).  Note that in doing so by use of Eq.(\ref{eq:ME}),
we make a somewhat arbitary choice, that the eigenvectors needed in Eq.(\ref{eq:weight}) are the mass-weighted displacements
$|s\rangle$, rather than the simple displacements $|u\rangle$.  This has an advantage of simplicity, but more than that,
the $|s\rangle$ states are orthonormal by virtue of the ordinary Hermitean nature of Eq.(\ref{eq:s}).  The states $|u\rangle$
have instead, the generalized orthonormality relations, $\langle u,\alpha|{\hat{\sf M}}|u,\beta\rangle = \delta_{\alpha,\beta}$.  
This would have altered
the sum rule on the weights.  Rather than adding to 1, they would add to a mass and $K$-dependent number.  
The answers for the simpler $|s\rangle$ eigenvectors is given in Eq.(\ref{eq:wph}) and Fig. \ref{fig:ph}.

\
\
\
\
\
\
\
\

\section*{Acknowledgements}
PBA thanks colleagues of SWaSSiT (Solar Water-Splitting Simulation Team).  We especially
thank M. V. Fern\'andez-Serra, J. Liu, and L. S. Pedroza for help.
PBA was supported in part by US DOE Grant No. DE-FG02-08ER46550.  
DAC thanks NSF for a summer REU scholarship administered by Stony Brook University (NSF PHYS-0851594). 
JMS was supported by grants FIS2009-12721 and FIS2012-37549
\
\
\
\
\
\
\
\
\
\
\
\
\
\
\
\
\
\
\
\
\
\
\
\
\

\end{document}